\input{aipcheck}

\documentclass[
    ,final            % use final for the camera ready runs
sort&compress
  ]
  {aipproc}

\layoutstyle{6x9}

\usepackage[para]{threeparttable}

\usepackage[]{aas_macros}

\begin{document}

\title{The Robotic Super-LOTIS Telescope: Results \& Future Plans}


\author{G.~G.~Williams}{
  address={MMT Observatory, P.O. Box 210065, University of Arizona, Tucson, AZ 85721-0065},
  email={gwilliams@as.arizona.edu}
}

\author{P.~A.~Milne}{
  address={Steward Observatory, 933 N Cherry Ave, Tucson, AZ, 85721}
}

\author{H.~S.~Park}{
  address={Lawrence Livermore National Laboratory, 7000 East Avenue, Livermore, CA 94550}
}

\author{S.~D.~Barthelmy}{
  address={NASA Goddard Space Flight Center, Greenbelt, MD 20771}
}

\author{D.~H.~Hartmann}{
  address={Department of Physics and Astronomy, Clemson University, Clemson, SC 29634}
}

\author{A.~Updike}{
  address={Department of Physics and Astronomy, Clemson University, Clemson, SC 29634}
}

\author{K.~Hurley}{
  address={Space Sciences Laboratory, University of California, Berkeley, CA 94720}
}

\begin{abstract}
We provide an overview of the robotic Super-LOTIS (Livermore Optical
Transient Imaging System) telescope and present results from
gamma-ray burst (GRB) afterglow observations using Super-LOTIS and
other Steward Observatory telescopes. The 0.6-m Super-LOTIS telescope
is a fully robotic system dedicated to the measurement of prompt and
early time optical emission from GRBs.  The system began routine
operations from its Steward Observatory site atop Kitt Peak in April
2000 and currently operates every clear night.  The telescope is
instrumented with an optical CCD camera and a four position filter
wheel. It is capable of observing Swift Burst Alert Telescope (BAT)
error boxes as early or earlier than the Swift UV/Optical Telescope
(UVOT).  Super-LOTIS complements the UVOT observations by providing
early R- and I-band imaging.  We also use the suite of
Steward Observatory telescopes including the 1.6-m Kuiper, the
2.3-m Bok, the 6.5-m MMT, and the 8.4-m Large Binocular Telescope
to perform follow-up optical and near infrared observations of
GRB afterglows.  These follow-up observations have traditionally 
required human intervention but we are currently working to
automate the 1.6-m Kuiper telescope to minimize its response time.
\end{abstract}

\maketitle

%%%%%%%%%%%%%%%%%%%%%%%%%%%%%%%%%%%%%%%%%%%%
%% MAINMATTER
%%%%%%%%%%%%%%%%%%%%%%%%%%%%%%%%%%%%%%%%%%%%

\section{Introduction}

The Super-LOTIS collaboration aims to obtain two types of
data sets that can be used independently or together to
provide a more complete understanding of the physics of GRBs.
These data sets are: (1)~very early optical imaging, and
(2)~follow-up optical and near infrared imaging, spectroscopy,
and polarimetry.  The early optical imaging is provided by a
fully robotic 0.6-m telescope and the follow-up data comes
from larger aperture Steward Observatory telescopes.

Early-time GRB optical counterpart identification is important
because it provides information such as: (1)~the peak optical flux,
useful for planning follow-up observations; (2)~the early decay
indices, useful in identifying subsequent flaring activity or breaks
in the light curve; (3)~color, useful in estimating redshift
and planning follow-up observations; and (4)~variability, useful
in understanding continued activity in the central engine or
structure in the circumburster medium \citep[e.~g.~][]{butler06}.
Perhaps the most scientifically valuable of these is the
variability which manifests itself in different forms that are
still not fully explained.

Follow-up imaging contributes to the long term light curves and
yields information about chromatic (spectral evolution) or
achromatic (jet) breaks.  In addition to providing a redshift,
spectroscopy can yield information about the circumburster
medium, the intergalactic medium, and in some cases the the
emerging supernova.  Finally, polarimetry is a powerful tool
for probing the nature of the jet and its evolution.

\section{System Overview}

Super-LOTIS is a fully robotic 0.6-m telescope dedicated to the
search for optical counterparts of GRBs.  The telescope is housed
in a roll-of\/f-roof facility at Steward Observatory's Kitt Peak site.
From 1999 to 2003 Super-LOTIS was configured with a prime focus
imager in order to provide better coverage of the large 
Burst and Transient Source Experiment (BATSE) error boxes.
Selected results from that system and its predecessor, LOTIS, are
included in \citet{park97, williams99, schaefer99, park02,
hurley02, castro02}, and \citet{blake03}.

The upgraded f/9 Cassegrain system uses a commercial Spectral
Instruments 800 Series CCD camera equipped with a thinned
$2048 \times 2048$ pixel E2V detector. The pixel scale of
$0\farcs5$/pixel is well matched to the image quality of the
optical design and the typical site seeing.  The
$17 \arcmin \times 17 \arcmin$ field-of-view provides full
coverage of the Swift BAT error boxes.  The system is linked
to the GCN network and can slew at a rapid rate of approximately
$8\arcdeg$/s which allows the telescope to begin imaging any part
of the sky within 25~s of receiving a GCN trigger.  Depending on
the observing conditions, the system can often achieve a limiting
magnitude of approximately $\rm{R} = 17.5$ in a single 10~s
exposure and $\rm{R} = 18.5$ in a single 60~s exposure.  When not
observing GRBs, the system performs ancillary observing
programs including nightly multi-band observations of supernovae
and searches for novae in M31 and M33.

The Super-LOTIS website, \url{http://slotis.kpno.noao.edu/LOTIS/index.php},
provides additional details, status information, and up-to-date results.
The characteristics of the system are provided in Table~\ref{tab:a}. 

\begin{table}
\begin{tabular}{lr}
\hline
Site: & Steward Observatory, Kitt Peak\\
Aperture: & 0.6-m\\
Instrument: & Spectral Instruments 800 Series Camera \\
Detector: & E2V TE Cooled CCD\\
Format: & $2048 \times 2048$, 13.5 $\micron$ pixels\\
Pixel Scale: & 0.5$\arcsec$/pix\\
Field-of-View: & $17\arcmin \times 17\arcmin$\\
Filters: & V, R, I, H-$\alpha$\\
Limiting Magnitude: & R $\sim$ 17.5 (10 s), R $\sim$ 18.5 (60 s)\\
Slew Speed: & $8\arcdeg$/s\\
Response Time: & < 25 s\\
Software: & Perl Client/Server\\
\hline
\end{tabular}
\caption{Super-LOTIS characteristics.}
\label{tab:a}
\end{table}

\section{Recent Results}

Table~\ref{results} provides a list of Swift era observations
obtained by our group which resulted in GCN Circulars\footnote{The
LBT data were obtained by P.~Garnavich and X. Dai as part of the
LBT Science Demonstration Time, see \citet{dai07}}.  The table is
not a complete log since some of our observations did not result
in GCN Circulars.  The upper section of Table~\ref{results} lists
events that were observed by Super-LOTIS and the lower section
includes observations using other Steward Observatory facilities.

Super-LOTIS has provided the earliest \emph{filtered} observations
of all eleven bursts that the system responded to promptly.
Several of those observations resulted in detections but even the
early-deep upper limits provide scientifically interesting constraints.
For events that were not observed promptly, either because of timing,
position, weather or a delayed trigger, deeper detections or limits
were achieved through co-addition of many individual frames.

Since the Swift UVOT is equipped with only blue filters the
Super-LOTIS R- \& I-band data complement the Swift UVOT data.
The color information can provide crucial clues about the
burst environment and early indications of the redshift.
In addition, the Super-LOTIS observations occasionally cover
gaps in the UVOT coverage which result from pointing or slewing
restrictions or periods of Earth blockage.  Optical observations
are critical during those gaps since flares or breaks in the
light curve could occur during them.

\begin{table}[htb]
%\begin{center}
\begin{threeparttable}[b]
\caption{\label{results}Recent observations resulting in GCN circulars.}
\begin{tabular}{|l|c|c|c|c|c|}
        \hline
UT Date &  Response Time & Exp. (s) & R Magnitude & GCN Circ. \\
        \hline
        \hline
071025   & 314.8 s (95.3 s) & $10 \times 60$ (10) & $17.97 \pm 0.17$ ($> 17.3$) & 6995 \\
071011   & 40.4 s & 10 & $> 16.9$ & 6887 \\
071010b   & 13.6 h & $59 \times 60$ & $> 20.4$ & 6893 \\
070612a   & 25.1 h & $6 \times 60$ & $17.73 \pm 0.4$ & 6535 \\
070610   & 2.5 d & $60 \times 60$ & $> 20.9$ & 6536 \\
070419a   & 456 s (102 s) & 60 (10) & $18.49 \pm 0.2$ ($> 17.7$) & 6328 \\
061126   & 35 s & 10 & $12.9 \pm 0.2$ & 5869 \\
061102b   & 38 s & $5 \times 10$ & $> 19.8$ & 5780 \\
061009   & 3.1 h & $14 \times 60$ & $> 20.2$ & 5742 \\
060923   & 41.1 s & 10 & $> 16.6$ & 5588 \\
060515   & 1745 s & $8 \times 60$ & $> 18.1$ & 5136 \\
060510b  & 1877 s & 60 & $> 18.6$ & 5100 \\
060502   & 1284 s (33 s) & 60 & $> 17.9$ & 5049 \\
060501   & 1.25 h & 60 & $> 18.6$ & 5045 \\
060210   & 94.5 s (55.1 s) & $5 \times 10$ (10) & $18.25 \pm 0.29$ ($> 17.5$) & 4730 \\
060206   & 4.2 h & $20 \times 60$ & $17.87 \pm 0.12$ & 4699 \\
060110   & 38.0 s & 10 & $> 16.0$ & 4469 \\
051111   & 35.9 s & 10 & $13.2 \pm 0.1$ & 4252 \\
051109b   & 56.0 s & 10 & $> 17.0$ & 4225 \\
051109a   & 43.0 s & 10 & $15.27 \pm 0.13$ & 4218 \\
050525   & 6.0 h & $20 \times 30$ & $> 17.5$ & 3485 \\
050421   & 3.8 h (8.0 h) & $30 \times 30$ & $> 17.5$ ($20.0$) & 3311 \\
\hline
\hline
071025   & 28 m & 150 & $18.7 \pm 0.1$ & 7011\tnote{1} \\
070419a   & 30.8 d & $15 \times 200$ & $25.71 \pm 0.13$ & 6486\tnote{2} \\
070419a   & 20.8 d & $25 \times 200$ & $25.29 \pm 0.5$ & 6486\tnote{2} \\
070419a   & 27 m & 60 & $19.12 \pm 0.09$ & 6341\tnote{1} \\
070125   & 26.8 d & $10 \times 200$ & $26.3 \pm 0.3$ & 6165\tnote{2} \\
060512   & 6.6 h & $5 \times 300$ & $20.14 \pm 0.16$ & 5127\tnote{1} \\
060121   & 5.6 h & $6 \times 300$ & $23.79 \pm 0.19$ & 4558\tnote{3} \\
051221b   & 6.5 h & $11 \times 300$ & $> 23.0$ & 4420\tnote{1} \\
050408   & 12.0 h & $11 \times 120$ & $21.9 \pm 0.1$ & 3258\tnote{1} \\
041217   & 19.5 h & $2 \times 300$ & $> 21.0$ & 2857\tnote{3} \\
        \hline
\end{tabular}
   \begin{tablenotes}
        % normally use the \tnote{1} command with these.
    \item [1] 1.6-m  Kuiper
    \item [2] 8.4-m LBT
    \item [3] 2.3-m Bok
   \end{tablenotes}
  \end{threeparttable}
%\end{center}
\end{table}

%%%%%%%%%%%%%%%%%%%%%%%%%%%%%%%%%%%%%%%%%%%%
%% Sample figure:
%%
%% The option [height=...] scales the picture to the given height,
%% without it it would be printed at its nominal size
%%%%%%%%%%%%%%%%%%%%%%%%%%%%%%%%%%%%%%%%%%%%

%\begin{figure}
%  \includegraphics[height=.3\textheight]{golfer}
%  \caption{Picture to fixed height}
%\end{figure}

%%%%%%%%%%%%%%%%%%%%%%%%%%%%%%%%%%%%%%%%%%%%
%% SAMPLE TABLE
%%
%% Shows the use of \tablehead and \tablenote
%% macros
%%%%%%%%%%%%%%%%%%%%%%%%%%%%%%%%%%%%%%%%%%%%

%\begin{table}
%\begin{tabular}{lrrrr}
%\hline
%  & \tablehead{1}{r}{b}{Single\\outlet}
%  & \tablehead{1}{r}{b}{Small\tablenote{2-9 retail outlets}\\multiple}
%  & \tablehead{1}{r}{b}{Large\\multiple}
%  & \tablehead{1}{r}{b}{Total}   \\
%\hline
%1982 & 98 & 129 & 620    & 847\\
%1987 & 138 & 176 & 1000  & 1314\\
%1991 & 173 & 248 & 1230  & 1651\\
%1998\tablenote{predicted} & 200 & 300 & 1500  & 2000\\
%\hline
%\end{tabular}
%\caption{Average turnover per shop: by type
%  of retail organisation}
%\label{tab:a}
%\end{table}

\section{Future Plans}

We plan to continue to operate Super-LOTIS as long as funding 
permits while at the same time expanding our use of other
Steward Observatory telescopes for GRB afterglow observations.
Toward this end, we have started the process of automating the
1.6-m Kuiper telescope.  The Super-LOTIS operations
software was written to be easily ported to the other Steward
telescopes.  We successfully demonstrated queue mode
operation of the Kuiper telescope in October 2007.  
We are currently assessing the hardware modifications
required to remotely control focus, mirror covers, and
the dome slit.  We anticipate that robotic or human assisted
automated operation will be available before the end of 2008.

The twin 8.4-m Large Binocular Telescope (LBT) will begin regular
science observing in February 2008.  The red and blue channel
prime focus cameras for the LBT can detect very faint sources
and therefore we plan to use them to obtain very late images to
search for supernova bumps, breaks in the light curve, and
host galaxies.

%%%%%%%%%%%%%%%%%%%%%%%%%%%%%%%%%%%%%%%%%%%%%%%%
%% BACKMATTER
%%%%%%%%%%%%%%%%%%%%%%%%%%%%%%%%%%%%%%%%%%%%%%%%

\begin{theacknowledgments}
This work was supported in part by NASA Proposal Number 06-SWIFT306-0067.
\end{theacknowledgments}



\bibliographystyle{aipproc}   % if natbib is available

\bibliography{ms}

\IfFileExists{\jobname.bbl}{}
 {\typeout{}
  \typeout{******************************************}
  \typeout{** Please run "bibtex \jobname" to optain}
  \typeout{** the bibliography and then re-run LaTeX}
  \typeout{** twice to fix the references!}
  \typeout{******************************************}
  \typeout{}
 }

\end{document}